\documentclass[journal]{IEEEtran}
\usepackage{amsmath,amssymb,amsfonts}
\usepackage{graphicx}
\graphicspath{{./}{figures/}{../6DMA_ISAC/}{../画图/}}
\usepackage{cite}
\usepackage{booktabs}
\usepackage{multirow}
\usepackage{bm}
\usepackage{tabularx}
\usepackage[table]{xcolor}
\usepackage{arydshln}
\usepackage[colorlinks,
linkcolor=blue,
anchorcolor=blue,
citecolor=blue]{hyperref}

\definecolor{TableBlue}{RGB}{225,238,248}
\definecolor{TableBlueLight}{RGB}{244,248,252}
\definecolor{TableTextBlue}{RGB}{31,78,121}
\definecolor{CommonBenefitColor}{RGB}{31,78,121}
\definecolor{IndividualBenefitColor}{RGB}{128,60,128}
\begin{document}

\title{\huge Reconfigurable Antenna-Enhanced ISAC: A Unified Perspective on Electromagnetic Adaptation}

\author{Peilan Wang,~\IEEEmembership{Member,~IEEE}, Lei Xie,~\IEEEmembership{Member,~IEEE}, Weidong Mei,~\IEEEmembership{Member,~IEEE},\\ Jun Fang,~\IEEEmembership{Senior Member,~IEEE}, Yihong Su,~\IEEEmembership{Member,~IEEE}, and Hongbin Li,~\IEEEmembership{Fellow,~IEEE}
\thanks{Peilan Wang, Weidong Mei, and Jun Fang are with National Key
Laboratory of Wireless Communications, University of
Electronic Science and Technology of China, Chengdu 611731, China
(email: peilan\_wangle@uestc.edu.cn; wmei@uestc.edu.cn; JunFang@uestc.edu.cn)}
\thanks{Lei Xie is with the School of Cyber Science and Engineering, Southeast University, Nanjing 211189, China (email: leixie@seu.edu.cn)}
\thanks{Yihong Su is with EHF Key Laboratory of Science, School of Electronic Science and Engineering, University of Electronic Science and Technology of China, Chengdu, 611731, China. Yihong Su is also with the Institute for Wireless Intelligence, Chengdu, 611730, Sichuan, China (email: yhsu@uestc.edu.cn)}
\thanks{Hongbin Li is with the Department of Electrical and Computer Engineering,
Stevens Institute of Technology, Hoboken, NJ 07030, USA (email:
Hongbin.Li@stevens.edu)}
}

\maketitle

\begin{abstract}
Integrated sensing and communication (ISAC) enables sensing and communication (S\&C) functions to share time, frequency, and hardware resources, offering an efficient paradigm for next-generation wireless systems. However, conventional fixed-configuration antenna (FCA) architectures provide limited flexibility in adapting the physical electromagnetic (EM) response of the end-to-end wireless link. This article develops a unified EM-adaptation perspective on reconfigurable antenna (RA)-enhanced ISAC systems. We distinguish transceiver-side (TSc-side) and environment-side (Env-side) RA deployments and identify five fundamental EM reconfiguration dimensions: position, orientation, radiation pattern, polarization, and frequency response. Building on this framework, we examine how these dimensions customize communication channels, actively control target information mapping for sensing, and reshape the interplay between S\&C, particularly by improving compatibility between S\&C requirements under shared resources. Two case studies illustrate how TSc-side and Env-side EM reconfiguration can improve spatial selectivity and the achievable S\&C performance tradeoff. Finally, we discuss several promising directions for advancing RA-enhanced ISAC systems, including fundamental performance limits, low-overhead black-box optimization, and artificial intelligence (AI)-enabled semantic ISAC.
\end{abstract}

\section{Introduction}

\begin{figure*}[!t]
    \centering
    \includegraphics[width=\textwidth]{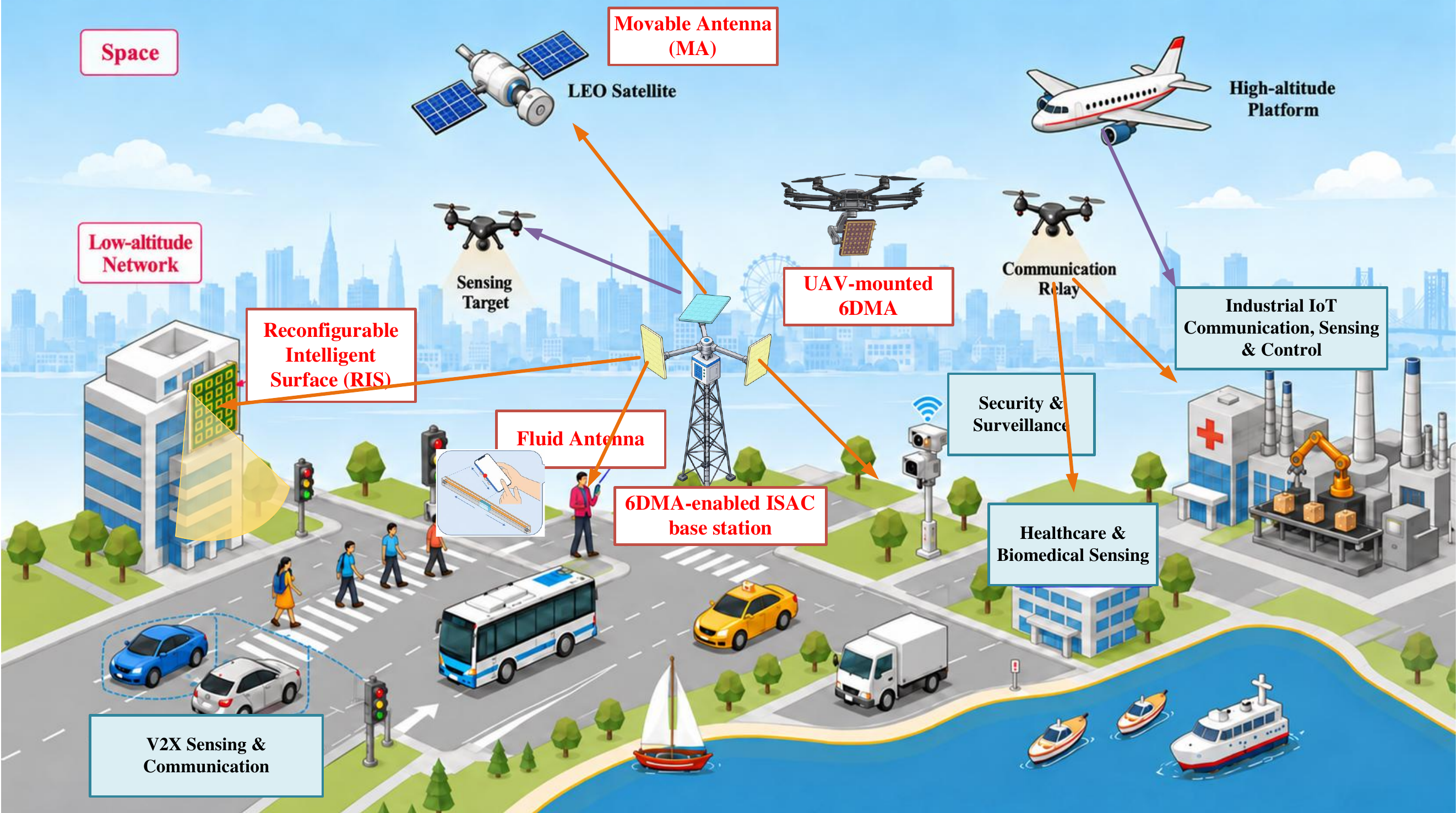}
     \caption{Illustration of RA-enhanced ISAC systems, where TSc-side RAs and Env-side RAs introduce additional EM DoFs to enhance S\&C performance across various scenarios.}
     \label{fig:system_fig}
\end{figure*}

Integrated sensing and communication (ISAC) aims to integrate two traditionally distinct functions, i.e., wireless sensing and communication (S\&C), over shared time, frequency, waveform, and hardware resources, thereby reducing redundant infrastructure and improving spectral efficiency \cite{LuLiu24}. This resource sharing supports applications such as autonomous driving, industrial automation, environmental monitoring, and low-altitude networks, making ISAC a key technology for next-generation wireless systems envisioned in International Mobile Telecommunications (IMT)-2030. However, S\&C impose different design requirements: sensing requires effective target illumination and echo collection, whereas communication requires reliable data delivery to users. Under shared resource constraints, these requirements may conflict with each other, leading to an inherent S\&C performance tradeoff.

Most conventional ISAC systems employ waveform design, beamforming, and receiver processing to manage this tradeoff, as increasing the array aperture or the number of antennas can improve spatial multiplexing capabilities for communication and angular resolution for sensing. Nevertheless, simply scaling the transceiver array may not guarantee commensurate performance gains. Since the available spatial degrees of freedom (DoFs) are constrained by the physical aperture and propagation environment, increasing antenna density within a fixed aperture does not necessarily introduce additional independent spatial modes or yield corresponding gains in sensing resolution and estimation accuracy. Moreover, deploying additional active elements and radio-frequency (RF) chains increases circuit complexity, power consumption, and hardware cost, particularly in high-frequency bands. More fundamentally, signal-level optimization with fixed antenna configurations can only adapt transmission and reception to the given channels, without the ability to alter the antenna geometry or the electromagnetic (EM) responses of fixed propagation structures. A central challenge is therefore to reshape end-to-end EM responses to simultaneously benefit S\&C.

As illustrated in Fig.~\ref{fig:system_fig}, reconfigurable antennas (RAs) have recently attracted increasing attention in wireless communications and sensing by introducing configurable physical and EM DoFs at transceivers and in the propagation environment \cite{zhu2026npj,MaSurvey26,zhao2026ra_survey}. Compared with conventional fixed-configuration antennas (FCAs), RAs can adjust their positions, orientations, radiation patterns, polarizations, and/or frequency responses on the transceiver side, thereby tuning the EM characteristics of the transmit and receive apertures. For example, movable antennas (MAs) and fluid antennas (FAs) exploit spatial channel variations by selecting favorable antenna positions within a confined region \cite{zhu2026npj,ning2025movable,FluidNew}, while six-dimensional movable antennas (6DMAs) further introduce three-dimensional (3D) rotational DoFs \cite{6DMAShao}. On the environment side, RAs deployed as reconfigurable propagation nodes, such as intelligent reflecting surfaces (IRSs), also known as reconfigurable intelligent surfaces (RISs), customize wireless propagation by imposing adjustable scattering responses on impinging EM waves \cite{IRS22wei}. Joint transceiver- and environment-side reconfiguration is also feasible, as exemplified by MA-IRS systems that combine antenna position control with programmable reflection \cite{MAIRS26}. 
Existing surveys have reviewed RA architectures and their operating principles \cite{MaSurvey26,zhao2026ra_survey}. Building on these studies, a cross-architecture understanding of how different EM controls contribute to communication and sensing performance, and particularly to their interplay, merits further investigation.

To bridge this gap, this article develops a unified EM-adaptation perspective for RA-enhanced ISAC systems. First, we classify RAs according to their deployment locations into transceiver-side (TSc-side) RAs and environment-side (Env-side) RAs, and identify five fundamental EM reconfiguration dimensions, namely, position, orientation, radiation pattern, polarization, and frequency response. Second, we explain how these dimensions customize end-to-end communication channels, modify the mapping of target properties into sensing observations, and examine how reconfiguration can improve channel compatibility and facilitate mutual assistance between S\&C. Third, two case studies are presented to illustrate the operating principle of RAs based on an active 6DMA base station (BS) and a passive 6D-IRS mounted on an uncrewed aerial vehicle (UAV), respectively. Finally, open research directions on fundamental limits, low-overhead black-box optimization, multi-timescale control, and AI-enabled semantic ISAC are discussed.

\begin{figure*}[!t]
    \centering
    \includegraphics[width=\textwidth]{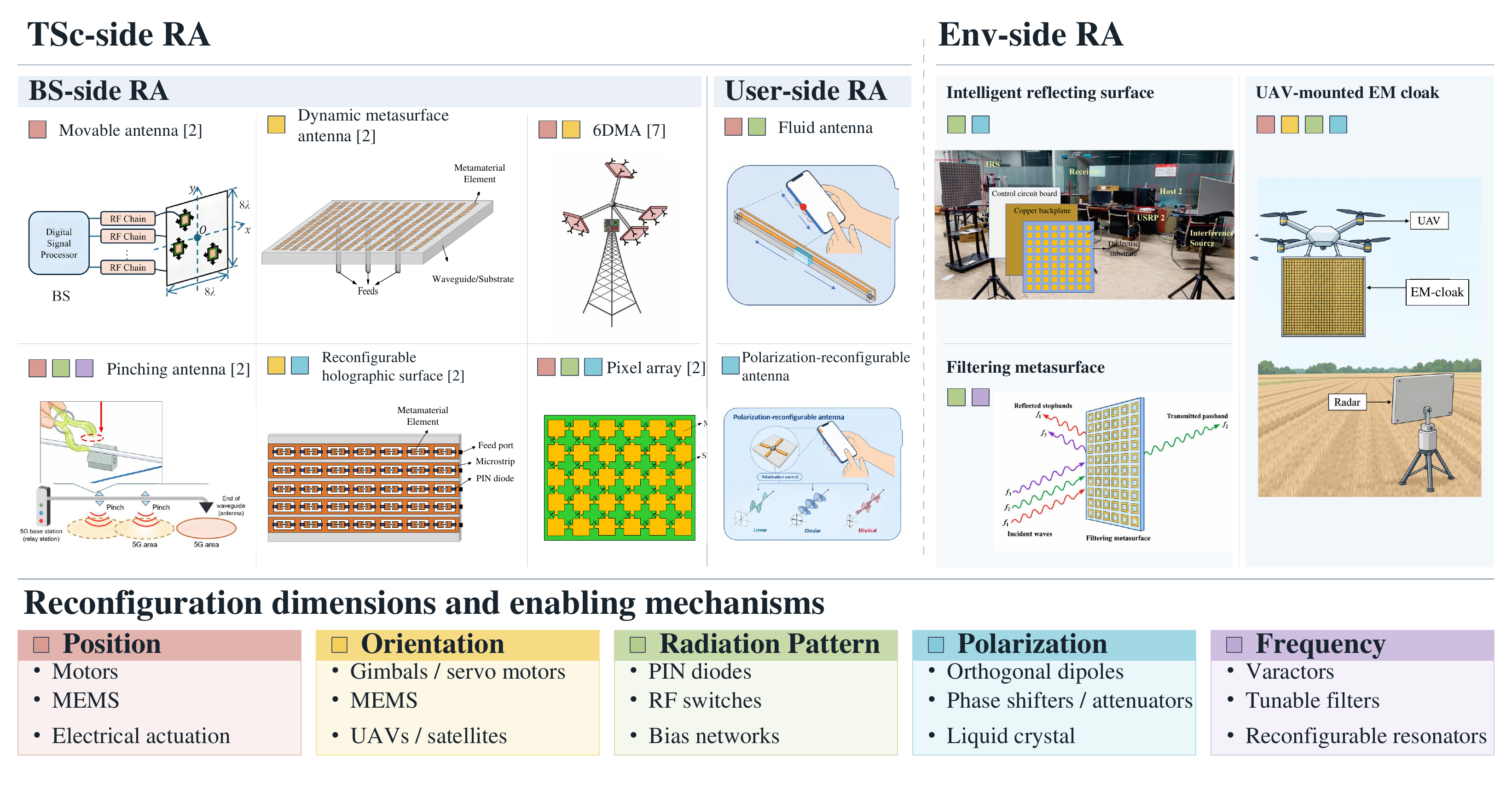}
     \caption{Deployment-based classification of RAs for ISAC. TSc-side RAs reconfigure transmit/receive apertures, whereas Env-side RAs adapt intermediate propagation paths. Their configurable EM properties comprise position, orientation, radiation pattern, polarization, and frequency response; multiple dimensions can be combined within one platform.}
    \label{fig:ra_taxonomy}
\end{figure*}

\section{Fundamentals of RA-Enhanced ISAC Systems}
This section establishes a unified view of RA-enhanced ISAC systems by classifying RAs deployed at the transceiver and within the propagation environment, and then introduces their five fundamental EM reconfiguration dimensions and representative enabling mechanisms.

\subsection{Rethinking RA-Enhanced ISAC Systems}

Although S\&C have different functional objectives, they share a common physical process of EM wave propagation: a transmitter radiates EM waves that interact with the environment before being collected by a receiver. In conventional FCA-based ISAC systems, transceiver antenna configurations and deployed propagation structures remain static after deployment, even though wireless channels vary dynamically due to user/target mobility and environmental changes. System adaptation consequently relies primarily on waveform design, beamforming, and receiver processing for the given physical configuration. In sharp contrast, RA-enhanced ISAC systems introduce controllable physical and EM DoFs into the transceiver apertures and the propagation environment, providing additional means to reshape the end-to-end S\&C channel responses.

As illustrated in Fig.~\ref{fig:ra_taxonomy}, we classify RAs according to their deployment locations into two complementary categories, namely \emph{TSc-side RAs} and \emph{Env-side RAs}. Specifically, TSc-side RAs refer to reconfigurable antennas or arrays deployed at the transmitter or receiver, whose physical or EM characteristics, including position, orientation, radiation pattern, polarization, and frequency response, can be dynamically adjusted. When multiple TSc-side RAs are integrated into an array, their joint configuration can further reshape the effective array geometry and spatial sampling aperture. On the other hand, Env-side RAs are deployed within the propagation environment between transceivers to modify the propagation paths for S\&C. Representative examples feature programmable metasurfaces (IRSs/RISs and their variants, such as simultaneously transmitting and reflecting RISs and beyond-diagonal RISs \cite{wu2024intelligent}), relays equipped with RAs, and reconfigurable backscatter nodes. Specifically, an IRS manipulates impinging EM waves through passive programmable scattering, an amplify-and-forward relay equipped with RAs adapts its receive and transmit aperture responses while retransmitting signals, and a reconfigurable backscatter node modulates incident RF waves to create additional useful propagation paths for symbiotic transmission.

These categories are complementary and can coexist within the same system. For instance, an Env-side relay can incorporate TSc-side RAs at its own transceivers. Through individual or joint configuration of TSc-side and Env-side RAs, the EM DoFs at both the transceivers and within the propagation environment can be tailored to meet diverse S\&C requirements \cite{zhu2026npj,zhao2026ra_survey}.

\subsection{Fundamental EM Reconfiguration Dimensions}

Both TSc-side and Env-side RAs manipulate EM waves by tuning one or more physical or EM properties, as summarized in Fig.~\ref{fig:ra_taxonomy}. To elucidate their operating principles, we classify these basic reconfiguration capabilities into five fundamental dimensions, namely, position, orientation, radiation pattern, polarization, and frequency response \cite{zhu2026npj,zhao2026ra_survey}. Note that these five dimensions are not necessarily independent, as adjusting one dimension may simultaneously affect another, and multiple dimensions can be jointly exploited in practical architectures to achieve higher flexibility.

\subsubsection{Position Reconfiguration}
Position reconfiguration changes the physical location of an antenna/surface or selects among spatially separated radiating ports. Practical implementations include motor-driven solid antennas, liquid-based antennas, and electronically switched ports \cite{ning2025movable,zhao2026ra_survey}. Note that mode switching may also shift the effective phase center, although it is not strictly equivalent to physical translation due to concurrent changes in the radiation pattern. Furthermore, mounting RAs on mobile platforms such as UAVs or ground vehicles extends position control to large-scale 3D space \cite{zhu2026npj,6DMAShao}.

\subsubsection{Orientation Reconfiguration}
Orientation reconfiguration physically rotates an antenna, subarray, or surface relative to the propagation directions \cite{6DMAShao,zhu2026npj}. Servo motors, rotational gimbals, and micro-electromechanical system (MEMS) actuators provide representative implementations \cite{6DMAShao,zhao2026ra_survey}. Physical rotation alters both the directional gain and the local polarization basis relative to the incident or outgoing wave, whereas electronic beam steering without physical rotation is categorized as radiation pattern reconfiguration, as discussed below.

\subsubsection{Radiation Pattern Reconfiguration}
Radiation pattern reconfiguration reshapes the angular radiation or scattering response by controlling surface current distributions or aperture fields. Representative implementations include switch-connected metallic pixels, feed switching, tunable parasitic radiators, programmable metasurface elements, and deformable conductors \cite{IRS22wei,zhao2026ra_survey}. Unlike rotating a fixed radiation pattern, these mechanisms can dynamically adjust beamwidths, sidelobe levels, and null directions without mechanical rotation.

\subsubsection{Polarization Reconfiguration}
Polarization reconfiguration adapts the polarization state of an antenna or scattering surface. Transceiver antennas can select or synthesize linear, circular, or elliptical polarization states by tuning excited modes, feeding networks, current paths, or the relative amplitudes and phases of orthogonal components \cite{MaSurvey26,zhao2026ra_survey}. For Env-side RAs, polarization-dependent scattering controls how incident polarization components are converted and coupled into reflected or transmitted waves.

\subsubsection{Frequency Reconfiguration}
Frequency reconfiguration shifts the operating band or shapes the frequency response of an aperture \cite{zhu2026npj,zhao2026ra_survey}. Modifying the effective electrical length, current paths, or reactive loading alters the antenna resonance or surface scattering characteristics, which is typically realized using PIN diodes, varactors, tunable materials, or reconfigurable resonators \cite{zhao2026ra_survey}. This physical EM-domain adaptation complements digital-domain carrier switching and subcarrier allocation.

\begin{table*}[!t]
\centering
\caption{EM effects and deployment-specific capabilities of the five reconfiguration dimensions. Each dimension lists shared effects (first line, blue) and illustrative deployment-specific capabilities (second line, purple).}
\label{tab:comparison}
{\footnotesize
\setlength{\tabcolsep}{4pt}
\renewcommand{\arraystretch}{1.18}
\begin{tabularx}{\textwidth}{>{\centering\arraybackslash}p{0.13\textwidth}XX}
\hline
\rowcolor{TableBlue}
\textbf{EM dimension} & \centering\arraybackslash\textbf{TSc-side RAs} & \centering\arraybackslash\textbf{Env-side RAs} \\
\hline
\textbf{Position} & \multicolumn{2}{c}{\textcolor{CommonBenefitColor}{\textbf{Channel-coupling adaptation and sensing-geometry control}}} \\
& \textcolor{IndividualBenefitColor}{Spatial selection diversity, adaptive spatial sampling, and virtual-aperture sensing} & \textcolor{IndividualBenefitColor}{Control of propagation geometry, coverage adaptation, and cascaded-link control} \\
\hline
\textbf{Orientation} & \multicolumn{2}{c}{\textcolor{CommonBenefitColor}{\textbf{Directional-coupling adaptation and angular selectivity}}} \\
& \textcolor{IndividualBenefitColor}{Adaptive transmit/receive aperture pointing} & \textcolor{IndividualBenefitColor}{Adaptive coupling between incident and scattered waves} \\
\hline
\textbf{Radiation pattern} & \multicolumn{2}{c}{\textcolor{CommonBenefitColor}{\textbf{Radiation/scattering pattern adaptation and spatial selectivity}}} \\
& \textcolor{IndividualBenefitColor}{Reconfigurable transmit/receive spatial filtering} & \textcolor{IndividualBenefitColor}{Programmable scattering and controllable non-LoS propagation} \\
\hline
\textbf{Polarization} & \multicolumn{2}{c}{\textcolor{CommonBenefitColor}{\textbf{Polarization matching and target/clutter discrimination}}} \\
& \textcolor{IndividualBenefitColor}{Polarization matching and, where supported, diversity/multiplexing} & \textcolor{IndividualBenefitColor}{Polarization-dependent propagation through tunable scattering responses} \\
\hline
\textbf{Frequency} & \multicolumn{2}{c}{\textcolor{CommonBenefitColor}{\textbf{Frequency-selective adaptation and spectral agility}}} \\
& \textcolor{IndividualBenefitColor}{Frequency-agile transmission/reception and multiband support} & \textcolor{IndividualBenefitColor}{Frequency-selective scattering and band-dependent path control} \\
\hline
\end{tabularx}
}
\end{table*}

These five dimensions are composable within a single hardware platform rather than being mutually exclusive categories \cite{zhu2026npj,zhao2026ra_survey}. For example, a 6DMA combines 3D position and 3D orientation controls \cite{6DMAShao}, while programmable pixel antennas and metasurfaces can integrate radiation pattern, polarization, and frequency controls. In practice, the selection of reconfiguration dimensions depends on deployment constraints, with BSs favoring scalable array architectures, user terminals subject to stringent size and power constraints, and Env-side RAs requiring cost-effective deployment and low-overhead coordination with transceivers.

\begin{figure*}[!t]
    \centering
    \includegraphics[width=\textwidth]{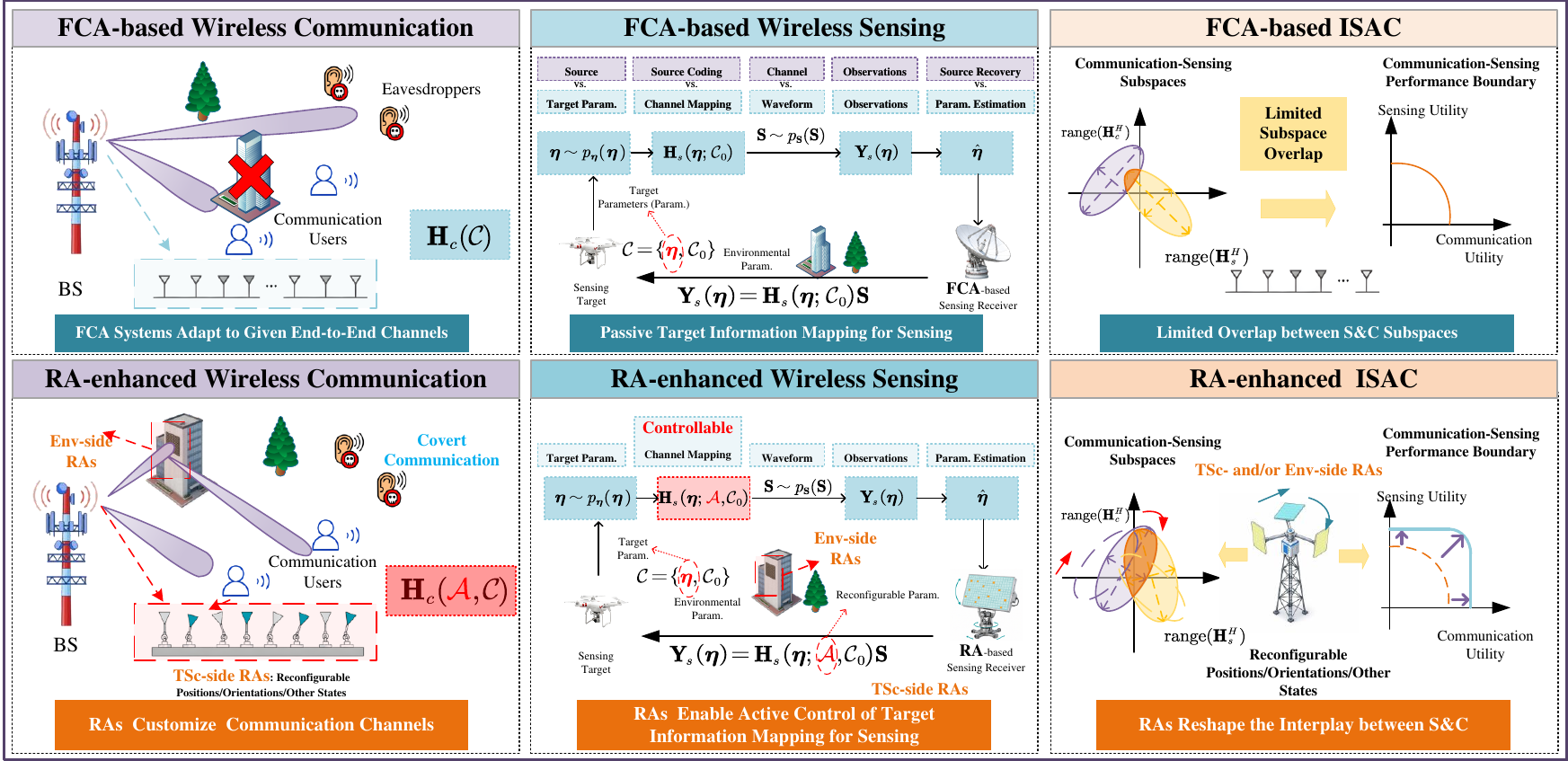}
     \caption{FCA-based and RA-enhanced communication, sensing, and ISAC. Left: adaptation of end-to-end communication channels. Middle: active control of target information mapping through configurable sensing responses. Right: improved compatibility between communication and target-illumination subspaces under shared transmit resources.}
     \label{fig:tradeoff}
\end{figure*}

\section{Benefits of RA-Enhanced ISAC Systems Through EM Adaptation}
\label{sec:benefits}

ISAC design seeks to jointly optimize S\&C performance under shared resource constraints, where the achievable performance region is governed by both individual S\&C capabilities and their mutual compatibility. Table~\ref{tab:comparison} summarizes the EM effects and deployment-specific capabilities of the five reconfiguration dimensions. Building on these mechanisms, Fig.~\ref{fig:tradeoff} illustrates how RAs customize communication channels, actively control target information mapping for sensing, and improve shared-resource compatibility. We discuss these individual benefits before examining integration gain and information-driven mutual assistance between S\&C.

\subsection{RAs Customize End-to-End Wireless Channels for Communications}

For wireless communications, RAs extend link adaptation from baseband signal processing to the physical EM responses underlying the end-to-end channel \cite{zhu2026npj,zhao2026ra_survey}. Despite diverse hardware implementations, the resulting communication benefits primarily manifest in two forms: \emph{channel enhancement} and \emph{channel discrimination}.

\emph{Channel enhancement} strengthens desired communication channels by improving aperture coupling and exploiting favorable propagation paths. TSc-side RAs adapt their aperture responses to dominant paths, while Env-side RAs create or reinforce useful propagation paths through programmable scattering \cite{IRS22wei}. These mechanisms can enhance channel gains and, when sufficiently independent spatial modes are available, improve multiple-input multiple-output (MIMO) channel conditioning. However, the achievable multiplexing gain remains geometry-dependent. For example, a single far-field line-of-sight (LoS) transmitter-IRS channel limits the cascaded channel to rank at most one, regardless of the configuration of the IRS phase shifts. Accordingly, higher spatial multiplexing gains require additional independent spatial modes, potentially enabled by richer scattering, distributed IRSs, or near-field propagation.
 
\emph{Channel discrimination} improves the relative channel strengths and spatial separability among different communication links. By adapting the coupling toward legitimate users, interferers, and potential eavesdroppers, EM reconfiguration can suppress undesired channels or reduce inter-user channel correlation \cite{xie2026secure,zhu2026npj}, thereby facilitating multiuser precoding, interference suppression, and secure transmission.

The five EM reconfiguration dimensions realize these common channel-level benefits through distinct physical mechanisms, while also providing dimension-specific advantages.
\begin{itemize}
    \item \emph{Position reconfiguration} exploits spatial channel variations by adjusting antenna locations to enhance desired links or mitigate interference. For example, MAs and FAs select favorable operating positions within a confined region, providing spatial selection diversity without increasing the number of simultaneously active antennas or RF chains \cite{ning2025movable,zhu2026npj}. Larger-scale repositioning can further alter propagation geometry and blockage conditions.
    \item \emph{Orientation reconfiguration} adjusts the coupling between an aperture and the surrounding propagation field. Rotating directional antennas or subarrays can improve alignment with favorable propagation paths while reducing coupling toward undesired directions \cite{6DMAShao,zhu2026npj}. Such flexibility also supports adaptive coverage and polarization alignment by changing the antenna orientation relative to the incident EM field.
    \item \emph{Radiation pattern reconfiguration} modifies the angular radiation response through tunable beamwidths, sidelobes, and nulls, enabling selective enhancement or suppression of propagation components \cite{zhao2026ra_survey}. It further provides flexible spatial coverage and complements conventional beamforming through additional control over the aperture response.
    \item \emph{Polarization reconfiguration} adapts the polarization response of an antenna by controlling the relative amplitudes
and phases of orthogonal electric-field components. Such adaptation improves polarization matching and enables selective coupling to signals with different polarization states \cite{MaSurvey26,zhao2026ra_survey}. It can further provide polarization diversity or multiplexing when supported by the propagation channel and transceiver hardware.
    \item \emph{Frequency reconfiguration} adapts the frequency-dependent radiation or scattering response to propagation conditions and spectrum requirements. This enables frequency-selective channel enhancement or interference suppression, while providing additional spectrum agility and multiband flexibility at the transceiver and environment sides \cite{zhao2026ra_survey}.
\end{itemize}

Collectively, the five EM reconfiguration dimensions provide complementary means of enhancing desired
communication channels and improving the discrimination among different links. Beyond these common channel-level
benefits, their distinct control capabilities offer additional advantages in spatial diversity, adaptive coverage, polarization diversity, and spectrum agility. The achievable gains depend on the propagation conditions and practical hardware constraints.

\subsection{RAs Actively Control Target Information Mapping for Sensing}

For sensing, a distinctive benefit of EM adaptation lies in
making the mapping from target parameters to received sensing
signals reconfigurable, as illustrated in Fig.~\ref{fig:tradeoff}.
Unlike communication, where information is deliberately encoded
 in transmitted symbols, sensing extracts target information
from propagation and scattering responses. Target parameters,
such as range, radial velocity, direction, and scattering
characteristics, are reflected in the delay, Doppler shift,
spatial response, and polarimetric signatures of the received
echoes, respectively.

Let $\mathcal{C}=(\boldsymbol{\eta},\mathcal{C}_0)$ denote the
propagation environment, where $\boldsymbol{\eta}$ collects
the target parameters and $\mathcal{C}_0$ represents the
remaining propagation and clutter parameters. For a fixed
antenna configuration, the mapping from the target parameters to the sensing channel, i.e.,
$\boldsymbol{\eta}\mapsto
\mathbf{H}_{\mathrm{s}}(\boldsymbol{\eta};\mathcal{C}_0)$, is determined by the antenna geometry and propagation
conditions. Introducing configurable RA states, collectively
denoted by $\mathcal{A}$, makes this mapping adjustable through
$\mathbf{H}_{\mathrm{s}}
(\boldsymbol{\eta};\mathcal{A},\mathcal{C}_0)$.
By adapting $\mathcal{A}$, the sensing system can modify
the target echo strength and the sensitivity of the sensing
response to target parameters without changing the target
state itself. The resulting sensing benefits depend on the waveform, noise,
 and clutter conditions. This controllable target information mapping
 provides additional flexibility for target detection,
parameter estimation, and target discrimination.

The five EM reconfiguration dimensions influence target
information mapping through distinct physical mechanisms,
while providing additional capabilities for specific
sensing tasks.

\begin{itemize}
    \item \emph{Position reconfiguration} adjusts the sensing
    geometry by changing antenna locations, propagation
    distances, and inter-element baselines. Such adjustments
    modify the array manifold and can improve the sensitivity
    of received signals to target directions and locations,
    potentially improving parameter estimation accuracy.
    Position selection can also provide complementary spatial
    observations, while larger-scale repositioning may reduce
    propagation losses or avoid blockage to improve target
    echo strength.

    \item \emph{Orientation reconfiguration} modifies the
    directional coupling between sensing apertures and
    target echoes. Rotating directional antennas or subarrays
    can align their high-gain directions with targets while
    reducing sensitivity toward clutter directions, thereby
    improving the target-to-clutter ratio. Different antenna
    orientations can also provide complementary directional
    responses for target localization and parameter
    estimation.

    \item \emph{Radiation pattern reconfiguration} adapts the radiation
    response through configurable beamwidths, sidelobes,
    and nulls. Narrow beams can concentrate illumination
    toward targets, while low-gain regions or nulls can
    suppress clutter and interference. Beyond these benefits,
    flexible beamwidth and beam-shape control allows sensing
    apertures to support different tasks, such as wide-area
    target search and focused target tracking.

    \item \emph{Polarization reconfiguration} adjusts the
    transmitted or received polarization states to exploit
    the polarimetric scattering characteristics of targets
    and clutter. Suitable polarization configurations can
    improve polarization matching for target echoes or suppress clutter
    with different polarization responses. Measurements
    acquired under multiple polarization states can further
    provide complementary polarimetric information for
    target discrimination and classification.

    \item \emph{Frequency reconfiguration} adapts the
    frequency-dependent radiation or scattering response
    to sensing requirements. Frequency-selective adaptation
    can improve target illumination or suppress undesired
    propagation components at selected frequencies.
    In addition, frequency-scanning antennas can provide
    angular scanning through frequency tuning, while
    multi-frequency observations can reveal complementary
    target scattering characteristics. When coherently
    processed over a sufficiently wide effective bandwidth,
    such observations can also improve range resolution.
\end{itemize}

Overall, EM reconfiguration provides additional control
over target information mapping by modifying the strength
and parameter sensitivity of received sensing responses.
These changes can improve target detection, parameter
estimation, and target discrimination, while the achievable
gains depend on the sensing task, propagation environment,
and available measurement resources.

\begin{figure*}[!t]
    \centering
    \includegraphics[width=\textwidth]{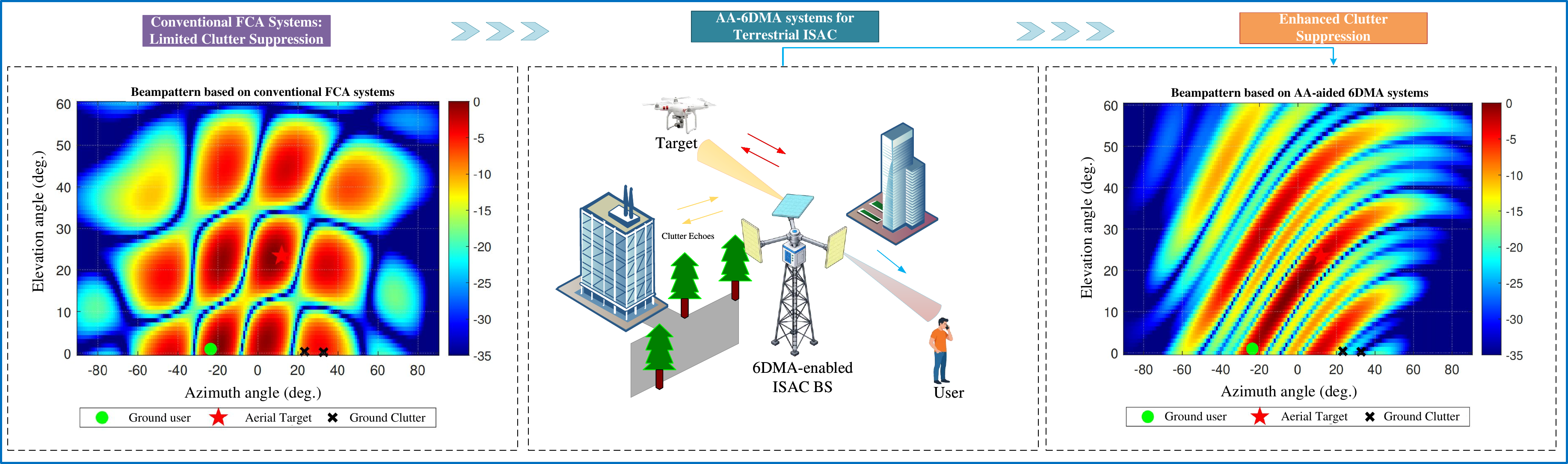}
     \caption{TSc-side AA-6DMA for clutter-resilient ISAC: conventional FCA transmit beampattern (left), system illustration (middle), and AA-6DMA transmit beampattern (right). }
    \label{fig:beampattern}
\end{figure*}

\begin{figure*}[!t]
    \centering
    \includegraphics[width=\textwidth]{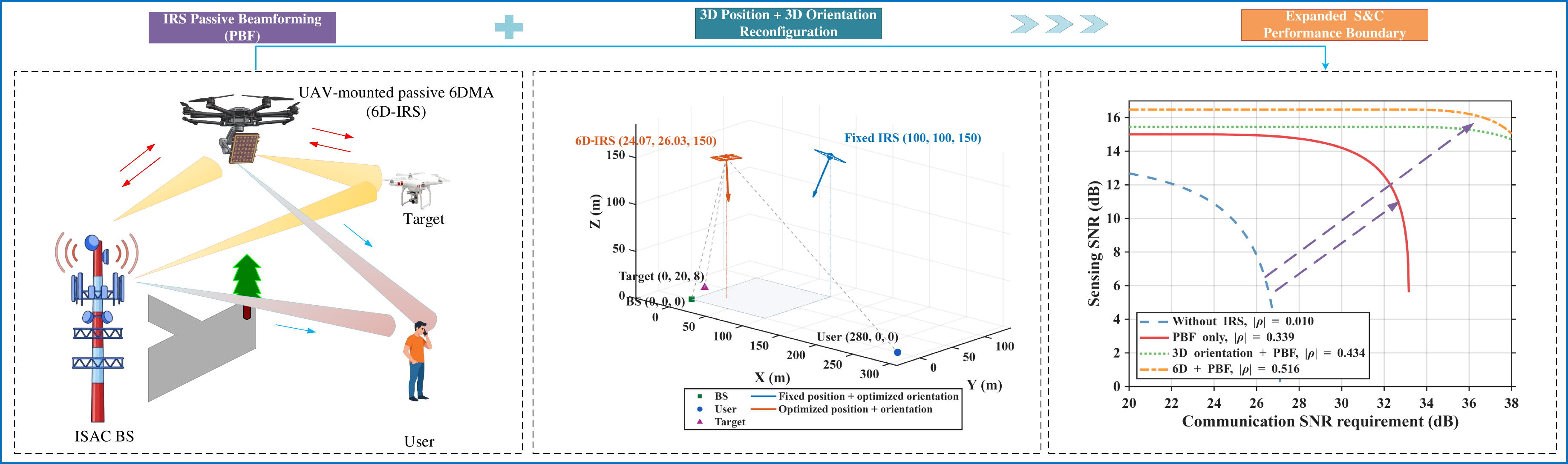}
     \caption{Env-side UAV-mounted 6D-IRS-assisted ISAC: system illustration (left), comparison of the geometric configurations of the UAV-mounted fixed-IRS and optimized 6D-IRS (middle), and achieved sensing SNR versus the minimum required communication SNR (right). PBF denotes passive beamforming, and $|\rho|$ is the normalized S\&C channel correlation magnitude. The IRS altitude is fixed at $150$~m, so only its horizontal position is optimized. The target is at $(0,20,8)$~m.}
     \label{fig:6dma_geometry}
\end{figure*}

\subsection{RAs Reshape the Interplay Between S\&C}
\label{sub:unified}

Beyond their individual benefits, RAs can further reshape
the interplay between S\&C through two complementary
mechanisms: \emph{integration gain} and \emph{coordination
gain}. The former improves the efficiency of shared
S\&C resources, whereas the latter exploits mutual
assistance between the two functions. In both
cases, RAs extend conventional signal-domain design
by making the physical apertures and propagation
environment reconfigurable.

\emph{Integration gain through channel compatibility:}
When S\&C share the same transmit waveform and resources,
the overlap between their effective channel subspaces
influences how efficiently the available spatial DoFs
can be jointly utilized. For example, greater overlap, which corresponds to a larger correlation coefficient between S\&C channel subspaces in \cite{LuLiu24,wang2025uav}, can enable a shared transmit beamformer to satisfy communication and target-illumination requirements more efficiently. Beyond subspace alignment, RAs can further improve
the utilization of shared resources by adapting different
EM responses to accommodate diverse S\&C requirements:
\begin{itemize}
    \item \emph{Position reconfiguration} modifies
    the array manifold and path-dependent phases,
    potentially improving the alignment between
    communication and target-illumination subspaces.
    The adjustable array geometry also provides
    flexibility to accommodate spatially separated
    users and sensing targets under shared
    beamforming constraints.

    \item \emph{Orientation reconfiguration} adjusts
    the directional coupling of antennas or subarrays
    toward communication users and sensing targets.
    By physically rotating the apertures, RAs can
    adapt their pointing directions to different
    angular distributions, particularly when users
    and targets occupy different elevation regions.

    \item \emph{Radiation pattern reconfiguration} adapts
    beamwidths, sidelobes, and radiation nulls
    to accommodate the spatial requirements of S\&C.
    In particular, configurable multi-lobe patterns
    can support simultaneous coverage of users
    and targets in different directions, even
    when their channel subspace overlap is limited.

    \item \emph{Polarization reconfiguration} adjusts
    the polarization responses to improve the
    compatibility of S\&C under a common polarization
    configuration. Alternatively, independently
    controllable polarization states can accommodate
    different communication and sensing requirements
    within a shared aperture, when supported by
    the transceiver hardware.

    \item \emph{Frequency reconfiguration} adapts
    the frequency-dependent radiation or scattering
    responses to different S\&C requirements.
    Besides selecting favorable common operating
    bands, it can support distinct frequency
    responses for S\&C within a shared multiband
    architecture.
\end{itemize}

On the environment side, RAs provide further
flexibility by introducing controllable propagation
paths that reshape the effective S\&C channels,
potentially improving their subspace compatibility
even under unfavorable direct-link geometry
\cite{wang2025uav}.
These mechanisms can improve shared-resource
utilization through subspace alignment and
complementary EM-domain controls.

\emph{Coordination gain through mutual assistance:}
Beyond integration, sensing-derived
user locations, motion, and blockage information
can guide the position, orientation, or pattern
reconfiguration of TSc-side and Env-side RAs
for communication beam tracking and blockage
avoidance. For example, sensing-based prediction
of user movement can enable proactive RA
reconfiguration before significant communication
channel degradation occurs. Conversely,
communication-derived channel or location
information may provide useful priors for
sensing-oriented RA configuration, potentially
reducing target search or measurement overhead
when relevant to the sensing task.
Such information-driven adaptation enables
mutual assistance between S\&C. Its effectiveness
also depends on the accuracy and timeliness of
the available information.

Integration and coordination gains offer
complementary means of improving joint S\&C
performance. While integration gain primarily
improves compatibility between S\&C requirements and the
efficiency of shared EM resource use, coordination gain exploits
cross-functional information to guide RA
adaptation. The following case studies mainly
illustrate how TSc-side and Env-side reconfiguration
improves the achievable S\&C performance tradeoff
through adaptive EM responses.

\section{Case Studies}
\label{sec:cases}
Two case studies illustrate TSc-side spatial selectivity and Env-side propagation control for joint S\&C performance.

\subsection{TSc-Side RA-Enhanced ISAC}

We consider a BS with four independently positionable and orientable 6DMA subarrays, each comprising a $2\times2$ uniform planar array at $28$~GHz ($16$ elements total). The BS serves a single-antenna ground user at $(-23.6^{\circ},0^{\circ})$ and tracks an aerial target at $(11.9^{\circ},24^{\circ})$, amid two strong ground clutter sources at $(23^{\circ},0^{\circ})$ and $(33^{\circ},0^{\circ})$; angle pairs denote azimuth and elevation. The assignment-aided 6DMA (AA-6DMA) scheme partitions subarrays between S\&C using geometric alignment scores. These scores measure how closely each subarray's outward normal aligns with the user or target direction, with better alignment yielding higher effective aperture gain. A weight $\alpha\in[0,1]$ balances communication rate and sensing mutual information. Each subarray translates within a $0.04$~m cube and rotates within $\pm35^{\circ}$ in elevation and $\pm65^{\circ}$ in azimuth relative to its assigned initial orientation.

At $\alpha=0.5$, Fig.~\ref{fig:beampattern} contrasts the FCA (left) and AA-6DMA (right) transmit beampatterns. Unlike the FCA's pronounced sidelobe responses toward clutter, AA-6DMA provides more selective responses toward the user and target with reduced radiation toward clutter. Orientation control improves directional coupling toward the aerial target, while position adjustments modify inter-subarray phase relationships that can help suppress clutter leakage, illustrating enhanced spatial selectivity.

\subsection{Env-Side RA-Enhanced ISAC}

We consider a $32$-antenna FCA BS assisted by a UAV-mounted $4\times4$ passive 6D-IRS at $3.6$~GHz with $20$~dBm BS transmit power \cite{wang2025uav}. The BS, user, and target are at $(0,0,0)$, $(280,0,0)$, and $(0,20,8)$~m, respectively, yielding weakly aligned direct S\&C channels. We compare ``Without IRS'', ``PBF only'' (IRS phase-shift optimization at fixed position $(100,100,150)$~m and orientation), ``3D orientation + PBF'' (joint orientation and phase-shift optimization at the same fixed position), and ``6D + PBF'' (joint horizontal-position, orientation, and phase-shift optimization). Horizontal motion is confined to a $100~\mathrm{m}\times100~\mathrm{m}$ region at a fixed altitude of $150$~m.

The middle panel of Fig.~\ref{fig:6dma_geometry} shows the IRS moving to $(24.07,26.03,150)$~m and tilting toward the BS--target region, reducing BS--IRS--target--BS echo path loss and improving aperture alignment. In the right panel, ``6D + PBF'' gains $6.71$~dB sensing SNR over ``Without IRS'' at a communication SNR requirement of $25$~dB. The gain reflects stronger channels and improved S\&C subspace alignment, with $|\rho|$ increasing from $0.010$ to $0.516$. PBF only shows a narrow flat range before rapid degradation in sensing SNR. Orientation optimization widens this range, while horizontal repositioning raises the tradeoff curve. Thus, Env-side reconfiguration improves the S\&C tradeoff by reshaping propagation paths and enhancing channel alignment.

\section{Future Directions}
\label{sec:challenges}
The additional DoFs offered by RAs bring performance gains but also raise new challenges in theory, optimization, and system control. In the following, we discuss several promising directions to address these challenges, including fundamental performance limits, low-overhead optimization, multi-timescale control, and artificial intelligence (AI)-enabled semantic ISAC.

\subsection{Fundamental Performance Limits}
A primary theoretical challenge is to characterize how the achievable S\&C performance region scales with the number of independent EM modes rather than the number of antenna elements or available configurations. For position reconfiguration, existing MA studies have established Cram\'{e}r-Rao bound (CRB) scaling laws with respect to observation duration and movement-region size~\cite{MAtrajectory}. Extending such bounds to general RAs requires quantifying the usable directional modes under constraints on 6DMA rotation and radiation patterns, the resolvable delay modes under frequency reconfiguration, and the polarimetric separability between targets and clutter. Because these five EM reconfiguration dimensions are physically coupled across finite apertures, a unified information-theoretic and estimation-theoretic framework is needed to characterize their joint performance limits imposed by channel rank, coherence time, and hardware constraints. Furthermore, since random communication symbols cause instantaneous sensing performance to fluctuate~\cite{xie2025sensing}, it is important to investigate whether dynamic RA configuration diversity can reduce these sensing-performance fluctuations without compromising the communication rate.

\subsection{Low-Overhead Black-Box Optimization}

Model-based optimization of RA configurations relies on channel state information (CSI) and accurate channel models, and acquiring the required CSI can incur substantial overhead \cite{Deriva26}.
For example, in MA- and 6DMA-enabled ISAC systems,
the effective S\&C channels vary with antenna positions
and orientations, making exhaustive channel acquisition
or configuration search impractical within the channel coherence time. Measurement-driven black-box optimization offers a promising
alternative by directly exploiting S\&C performance
feedback without requiring full CSI or analytic channel modeling. For instance,
Bayesian optimization can leverage limited measurements
to guide the selection of MA positions or 6DMA configurations,
potentially reducing the number of required configuration evaluations.
Communication feedback can often be obtained from pilot measurements, whereas sensing feedback may additionally require echo processing and clutter estimation. Developing measurement-efficient methods that account
for heterogeneous S\&C feedback and practical
reconfiguration constraints remains an important
research direction.

\subsection{Multi-Timescale Optimization and Hierarchical Control}

RA configurations and baseband processing operate on
different timescales due to their hardware constraints
and channel dynamics. Electronic reconfiguration of
radiation patterns, polarization states, and IRS phase
shifts is generally faster than mechanical adjustment
of MA positions and 6DMA orientations. However, the
appropriate update rates also depend on the sensitivity
of S\&C performance to different RA states and the
variations in propagation conditions. Updating all
parameters at the same rate may thus incur unnecessary
overhead or compromise the ability to adapt to fast channel variations. A promising direction is adaptive multi-timescale
control that jointly considers hardware response times
and S\&C requirements. For example, 6DMA positions
and orientations can be optimized over longer intervals
based on user and target distributions, while digital
beamforming and electronic RA states adapt to short-term
channel variations. For sensing, maintaining a stable
antenna geometry during coherent processing intervals
facilitates echo integration, whereas reconfiguration
between intervals can provide complementary spatial
observations. The key challenge is to determine which
RA states to update and when, considering their impact
on S\&C performance, channel dynamics, and
reconfiguration overhead.

\subsection{AI-Enabled Semantic ISAC with RAs}

In semantic ISAC, semantic sensing extracts task-relevant attributes of targets and the environment, while semantic communication conveys information needed for receiver-side tasks. The output of semantic sensing can also be delivered to communication users, as in cooperative
perception for connected vehicles.

RAs can support semantic ISAC by adapting their
EM configurations to task-specific sensing and
communication requirements. For example, a
6DMA-enabled BS can adjust its antenna geometry
to acquire informative target observations,
while Env-side RAs facilitate the delivery of
the resulting semantic information. AI-enabled
wireless world models can further integrate
multimodal sensing data and learned environmental
dynamics to predict task-relevant conditions
and guide proactive RA reconfiguration.
A promising direction is to jointly optimize
what information to sense, what information to
transmit, and how to configure RAs according
to task-level performance, while accounting
for environmental uncertainty and limited
measurement resources.

\section{Conclusion}
\label{sec:conclusion}
This article presented a unified EM-adaptation perspective on RA-enhanced ISAC systems covering TSc-side and Env-side architectures and five fundamental reconfiguration dimensions, namely position, orientation, radiation pattern, polarization, and frequency response. Through this framework, we explained how these EM controls customize end-to-end communication channels, enable active control of target information mapping for sensing, and reshape the S\&C interplay, particularly by improving compatibility between S\&C requirements under shared EM resources. Information-driven mutual assistance offers a complementary mechanism for S\&C coordination. Two case studies illustrated how TSc-side and Env-side reconfiguration can improve spatial selectivity and the achievable S\&C performance tradeoff. Finally, fundamental performance limits, low-overhead black-box optimization, multi-timescale control, and AI-enabled semantic ISAC were identified as promising future research directions. Overall, this perspective highlights the potential of jointly designing physical EM responses and signal processing to advance resource-efficient and adaptive ISAC systems.

\end{document}